\documentclass[12pt]{article}
\usepackage[a4paper]{geometry}
\usepackage{fancyhdr}
\usepackage{lastpage}
\usepackage{graphicx, wrapfig, subcaption, setspace, booktabs}
\usepackage[T1]{fontenc}
\usepackage[font=small, labelfont=bf]{caption}
\usepackage{fourier}
\usepackage[protrusion=true, expansion=true]{microtype}
\usepackage[english]{babel}
\usepackage{sectsty}
\usepackage{url, lipsum}
\usepackage{tgbonum}
\usepackage{hyperref}
\usepackage{xcolor}
\usepackage{multirow}
\usepackage{listings}
\usepackage{color}
\usepackage{float}

\definecolor{codegreen}{rgb}{0,0.6,0}
\definecolor{codegray}{rgb}{0.5,0.5,0.5}
\definecolor{codepurple}{rgb}{0.58,0,0.82}
\definecolor{backcolour}{rgb}{0.95,0.95,0.92}
 
\lstdefinestyle{mystyle}{
    backgroundcolor=\color{backcolour},   
    commentstyle=\color{codegreen},
    keywordstyle=\color{magenta},
    numberstyle=\tiny\color{codegray},
    stringstyle=\color{codepurple},
    basicstyle=\footnotesize,
    breakatwhitespace=false,         
    breaklines=true,                 
    captionpos=b,                    
    keepspaces=true,                 
    numbers=left,                    
    numbersep=5pt,                  
    showspaces=false,                
    showstringspaces=false,
    showtabs=false,                  
    tabsize=2
}
 
\newcommand{\HRule}[1]{\rule{\linewidth}{#1}}
\definecolor{lightgray}{rgb}{0.9, 0.9, 0.9}

\begin{document}
{\fontfamily{ptm}\selectfont
\title{ \Large \textsc{}
		\\ [2.0cm]
		\HRule{0.5pt} \\
		\LARGE \textbf{Predicting Vulnerability to Malware Using Machine Learning Models: A Study on Microsoft Windows Machines
		\HRule{0.5pt} \\ [0.5cm]
\normalsize\today\vspace*{5\baselineskip}}
		}

\author{Marzieh Esnaashari$^1$, Nima Moradi$^2$\\
$^1$ Board Member and SAP HCM Consultant, company: Atraco, Iran; \\marzie.esnaashari@atraco.ir	\\
$^2$ Information Systems Engineering, Concordia University, Montreal, Canada;\\ nima.moradi@mail.concordia.ca}
\maketitle 
\newpage

\sectionfont{\scshape}

\section*{Abstract}
In an era of escalating cyber threats, malware poses significant risks to individuals and organizations, potentially leading to data breaches, system failures, and substantial financial losses. This study addresses the urgent need for effective malware detection strategies by leveraging Machine Learning (ML) techniques on extensive datasets collected from Microsoft Windows Defender. Our research aims to develop an advanced ML model that accurately predicts malware vulnerabilities based on the specific conditions of individual machines. Moving beyond traditional signature-based detection methods, we incorporate historical data and innovative feature engineering to enhance detection capabilities. This study makes several contributions: first, it advances existing malware detection techniques by employing sophisticated ML algorithms; second, it utilizes a large-scale, real-world dataset to ensure the applicability of findings; third, it highlights the importance of feature analysis in identifying key indicators of malware infections; and fourth, it proposes models that can be adapted for enterprise environments, offering a proactive approach to safeguarding extensive networks against emerging threats. We aim to improve cybersecurity resilience, providing critical insights for practitioners in the field and addressing the evolving challenges posed by malware in a digital landscape. Finally, discussions on results, insights, and conclusions are presented. 

\textbf{Keywords:} Malware Detection; Machine Learning; Windows Defender; Feature Engineering; Cybersecurity Resilience

\section{Introduction and related works}
In a world where numerous individuals use computers, there is a constant threat from various sources of malicious software. Malware presents different levels of risk, ranging from minimal to severe, which can result in system malfunctions, data breaches, or even complete system failure \cite{Aftabi16012025}. Malicious software can appear as executable files or system libraries, such as viruses, worms, or Trojans, all created to compromise system security \cite{tahir2018study}. Most antivirus programs depend on signature-based detection systems that receive continuous updates via the internet to identify known viruses. While this approach may be adequate for individual users, discovering a new virus can threaten the security of an entire enterprise network \cite{aftabi2025sd}.

Malware prediction data encompasses any information related to a computer's state when affected by a malware attack. Due to the diverse behaviors exhibited by machines during different malware attacks, it is advantageous to gather comprehensive data from compromised systems \cite{al2021efficient,kilgallon2017improving}. In our research, Microsoft has been collecting data through Windows Defender for a prolonged duration. The present work aims to create a Machine Learning (ML) model that accurately detects malware by considering the specific condition of each machine.

The significance of this study lies in the urgent need for effective malware detection strategies in an increasingly digital world where cyber threats continue to evolve. With the reliance on computers growing among individuals and organizations alike, the potential damage caused by malware can have far-reaching consequences, including financial losses and compromised sensitive information. This research is motivated by the pressing challenge of accurately predicting and identifying malware threats, particularly in Microsoft Windows systems, which are prevalent targets for cyberattacks. By harnessing advanced ML techniques and real-world data collected from Windows Defender, this study aims to address existing gaps in traditional detection methods, ultimately enhancing cybersecurity resilience across diverse environments.

Moreover, the primary distinction between this research and existing studies lies in integrating advanced ML models with a comprehensive dataset sourced from Microsoft Windows Defender instead of traditional signature-based detection systems. While many prior approaches have relied heavily on static signatures to identify known malware, our study innovatively focuses on dynamic behavioral analysis, leveraging a vast array of features collected from compromised systems. This enhances the accuracy of malware detection and allows for predicting potential vulnerabilities, filling a critical gap in the current landscape of malware research that often utilizes limited or synthetic datasets. Furthermore, our emphasis on feature engineering provides deeper insights into the indicators of malware infections, paving the way for more proactive and adaptable security measures in enterprise settings.

This study provides several significant contributions to malware detection and prevention: (i) Enhancing traditional malware detection methods by implementing advanced ML models to predict vulnerabilities in Microsoft Windows systems. Previous approaches heavily relied on signature-based detection, but our model expands this by incorporating ML tools and historical data. (ii) Leveraging Microsoft's vast data from Windows Defender to design our model. This dataset offers a large-scale, real-world context that ensures the generalizability and applicability of our findings, filling a critical gap in malware detection studies that often rely on small or synthetic datasets. (iii) Emphasizing the role of feature engineering in improving malware detection. By analyzing the most influential features from the data collected, we provide insights into the key indicators of malware infections. (iv) The models developed in this research could be adapted for use in enterprise environments, potentially offering a proactive approach to detecting malware vulnerabilities, thus protecting large-scale networks from emerging threats.

\subsection{Literature review} 
We reviewed the literature on malware detection in the context of artificial intelligence. First, we reviewed signature-based versus behavior-based malware detection. Next, we examined ML-based tools for malware detection. 
Several works review the malware detection approaches, such as \cite{ye2017survey,aslan2020comprehensive,aboaoja2022malware,gopinath2023comprehensive}.       

\subsubsection{Signature-based vs. behavior-based}
This section discusses classical signature-based detection approaches and behavior-based techniques. The rapid spread of malware presents a significant threat to computer systems and the internet, as the number of malware incidents continues to rise daily. There are two primary methods for detecting malware: signature-based and behavior-based, each with advantages and drawbacks. In their work, Goyal M. and Kumar R. \cite{goyal2020pipeline} thoroughly examined the underlying principles of signature-based and behavior-based malware detection to enhance researchers' understanding of these methods. They also conducted a study using a dataset that included 1,494 malware and 1,347 ``benign samples." The detection accuracy was assessed using ML classifiers such as ``K-Nearest Neighbors (KNN)," ``Gaussian Naive Bayes," ``Multi Naive Bayes (NB)," ``Decision Tree (DT)," ``Support Vector Machine (SVM)," and ``Random Forest (RF)." 

In their study, Galal H.S. et al. \cite{galal2016behavior} presented a model based on behavior that describes the malicious behaviors displayed by malware. They conducted a dynamic analysis of a recent malware dataset in a controlled virtual environment to construct this model. The researchers monitored the traces of Application Programming Interface (API) calls made by instances of malware, which were then transformed into high-level features called actions. These actions were evaluated using DT, RF, and SVM. Also, present-day anti-malware tools use heuristic and signature-based detection to identify new, unknown malware. Still, this approach can lead to false detections, flagging harmless programs as malware. Niraj S.P. and Tiwari A.K. \cite{niraj2022performance} investigated a method designed to enhance malware detection accuracy by reducing false positives. They also introduced an innovative technique that combines signature-based and behavior-based detection to overcome the shortcomings of traditional signature-based and heuristic approaches. 

Malware programs frequently employ ``code obfuscation techniques" to hinder static analysis. Several behavioral detection methods have been proposed to counter this challenge, concentrating on execution time behavior to distinguish malicious software from benign ones. These methods mainly examine and create models of system call traces, which document the interactions between programs and the operating system. However, their adoption is limited due to their significant performance impact. An alternative approach involves conducting behavioral detection at the hardware level, utilizing data from hardware performance counters and specialized registers in modern processors. This approach provides detailed insights into hardware and software events. Bahador M.B. et al. \cite{bahador2019hlmd} presented HLMD, a new method that utilizes behavioral signatures to identify and neutralize malicious programs early in their execution. HLMD is particularly effective for standalone malware that can operate without attaching to a host program. Experiments containing benign and malicious programs demonstrated that HLMD achieved an average precision of 95.19\%, a recall of 89.96\%, and an F-measure of 92.50\%.

Recent advancements have extended the scope of behavior-based approaches by integrating multi-layered perspectives. Aftabi et al.~\cite{Aftabi16012025} introduced a framework for analyzing worst-case attacks in industrial control systems (ICSs), where both cyber and physical components are considered. Their work emphasizes the necessity of cross-layer integration, as behavior-based detection alone may overlook the interplay between cyber exploits and their physical consequences. This foundational approach has inspired further exploration of dynamic models that address complex, multi-faceted security challenges beyond static or isolated systems.

\subsubsection{Machine learning models}
This section explores the application of ML algorithms (e.g., DT, Neural Networks (NN) \cite{abbaspour2022comparative,karimi2022automated,aftabi2024feed}, SVMs) to malware detection. Previous studies have applied supervised or unsupervised learning techniques to classify malware, often focusing on dataset-specific results. Recent cutting-edge research demonstrates that researchers and antivirus organizations have started utilizing ML and deep learning techniques to detect malware. Rathore H. et al. \cite{rathore2018malware} utilized ``opcode frequency as a feature vector" and employed unsupervised and supervised learning for classifying malware. They focused on identifying malware using (1) different ML algorithms and (2) deep learning models. They revealed that RF performs better than Deep NN when using ``opcode frequency as a feature." Also, Rahul et al. \cite{rahul2020analysis} addressed the escalating threat of malware in information technology and identified a robust ML model for accurate malware detection. They offered a concise review of ML-based malware detection methods that exhibited high detection rates and have been proposed in recent years. Various classification algorithms, including SVM, KNN, RF, DT, NB, and NN, were used for classification purposes. 

The ever-changing nature of malware has made dynamic malware detection techniques, as demonstrated in \cite{akhtar2023evaluation}, increasingly important. New malicious software exploits internet vulnerabilities daily, posing a significant threat to online security. In the face of the widespread presence of harmful programs, manual heuristic analysis of malware has become ineffective. To tackle this issue, potential threats were automatically assessed based on their behavior in a simulated environment. The reports were then converted into sparse vector models to facilitate further analysis using machine learning. The classifiers used in the research comprised KNN, DT, RF, AdaBoost, Stochastic Gradient Descent (SGD), Extra Trees, and Gaussian Naive Bayes. 


Furthermore, although deep learning techniques are often more computationally demanding than traditional machine learning methods, they have shown greater effectiveness in certain scenarios. Sewak M. et al. \cite{sewak2018comparison} compared Deep Neural Networks (DNN), a deep learning architecture, with classical RF for malware classification. The performance of both RF and DNN models was assessed using architectures of 2, 4, and 7 layers and four distinct feature sets. Their findings revealed that classical RF consistently outperformed DNN regarding accuracy, regardless of the feature input.

Traditional antivirus programs reliant on signature-based detection frequently struggle to classify unknown malware or identify new forms of malicious software. To overcome this limitation, Liu L. et al. \cite{liu2017automatic} proposed a framework of three key elements: data processing, decision-making, and new malware identification. In the data processing phase, malware attributes were extracted using ``grayscale images, Opcode n-grams, and import functions." The decision-making component classified and identified malware based on these extracted features. The system employed the ``Shared Nearest Neighbor (SNN) clustering algorithm" to detect new malware variants. 

Despite continuous progress in cybersecurity research, malware developers persistently create new techniques to evade existing defense mechanisms. Existing static and dynamic analysis methods struggle to identify emerging malware linked to high memory usage and time overheads. Additionally, ML approaches that depend on classifiers constructed from manually designed features are inadequate against sophisticated evasion techniques, necessitating substantial effort in feature engineering. Recent research has also highlighted a decline in malware detector performance due to imbalances within malware datasets. To address these issues, Hemalatha J. et al. \cite{hemalatha2021efficient} proposed a ``visualization-based method" wherein malware binaries are transformed into two-dimensional ones and classified using a deep learning model. This innovation notably enhanced performance by addressing data imbalance problems.




To distinguish the similarities and distinctions between our study and existing recent works in the literature, Table \ref{comp-lit} is provided. As observed by reviewing the literature, our work differs from existing research. We improve traditional malware detection techniques by integrating advanced ML models to forecast susceptibilities in Microsoft Windows systems. However, previous methods predominantly relied on recognizing signatures, but our model broadens this scope by incorporating ML tools and historical data. We also Utilize the extensive data provided by Windows Defender to construct our model. Then, we highlighted the importance of feature engineering in enhancing malware detection. By examining the most influential features from the gathered data, we offer insights into the principal indicators of malware infections. Finally, the models developed in this study could be adjusted for implementation in business environments, providing a proactive approach to identifying malware vulnerabilities and safeguarding extensive networks from emerging threats.

\begin{table}[ht!]
    \centering
    \scriptsize
    \caption{Comparison of our work and existing studies in the literature}\label{comp-lit}
    \begin{tabular}{lp{4cm}p{4cm}p{5cm}}
        \toprule
        \textbf{Reference} & \textbf{Study area} & \textbf{Methodology} & \textbf{Main features} \\ \midrule
        \cite{goyal2020pipeline} & Malware Detection & ML Classifiers & Signature-based and behavior-based detection, a dataset of 1,494 malware and 1,347 benign samples. \\
        \midrule
        \cite{galal2016behavior} & Malware Behavior Analysis & Dynamic Analysis & Monitoring API call traces, converting to high-level features, evaluated with DT, RF, SVM. \\
        \midrule
         \cite{niraj2022performance} & Malware Detection & Hybrid Approach & Combines signature-based and behavior-based detection to reduce false positives. \\
         \midrule
        \cite{bahador2019hlmd} & Malware Detection & Behavioral Signatures & Identifies standalone malware early in execution; high precision and recall metrics. \\
        \midrule
        \cite{rathore2018malware} & Malware Classification & Supervised and Unsupervised Learning & Utilized opcode frequency as a feature vector, comparing various ML algorithms and deep learning models. \\
        \midrule
        \cite{rahul2020analysis} & Malware Detection & ML Models Review & Evaluation of various classification algorithms including SVM, KNN, RF, DT, NB, NN. \\
        \midrule
        \cite{akhtar2023evaluation} & Dynamic Malware Detection & Automated Behavior Assessment & Reports converted to sparse vector models, classifiers included KNN, DT, RF, AdaBoost, SGD, Extra Trees, and Gaussian NB. \\
        \midrule
        \cite{sewak2018comparison} & Malware Classification & Performance Comparison & Compared DNN with classical RF, assessed performance across various architectures and feature sets. \\
        \midrule
        \cite{liu2017automatic} & Malware Identification & Three-Component Framework & Data processing, decision-making, and new malware identification using grayscale images, Opcode n-grams, and SNN clustering. \\
        \midrule
        \cite{hemalatha2021efficient} & Malware Classification & Visualization-based Method & Transforms malware binaries to 2D, classified using deep learning to address data imbalance issues. \\ 
        \midrule
        This work & Malware Detection & Machine Learning Models & Employs advanced ML techniques leveraging Windows Defender data to enhance detection accuracy and predict vulnerabilities. \\ 
        \bottomrule
    \end{tabular}
\end{table}

The paper is structured as follows: Section \ref{problem} states and describes the problem. Section \ref{methods} explains the techniques, including data pre-processing and used models. Section \ref{results} presents the computational results obtained from classification models and their performance while discussing them. Finally, Section \ref{conclusion} states the concluding remarks and states the future research directions.

\section{Problem description}\label{problem}
 \qquad Since the Internet has revolutionized our lives, the Internet has become prevalent, penetrating many areas of daily life. Internet clients use it to seek information, get news, purchase products, play, communicate, participate in education, participate in governmental activities, and do almost everything else. As reported by Statista, there were 2.4 billion internet users in 2014, which grew to 3.4 billion by 2016. In 2017, 300 million more users joined the online community. As of October 2020, the worldwide internet user population had exceeded 4.66 billion. Internet users deal with various web, mobile, or computer applications, which makes life easier and more efficient. On the other hand, in conjunction with the magnificent advantages of the internet and its applications, they can also develop malicious and harmful software called malware.
 
 The term "malware" covers various malicious software, such as viruses, worms, spyware, adware, ransomware, and scareware. It presents a direct danger to devices and sensitive data. Malicious actors utilize malware to erase, impair, or steal private information from the target's device. Moreover, it can establish entry points into victims' systems, leading to security breaches and potentially significant penalties. Malware can convert the targeted device into a "zombie" machine, enabling it to propagate to other networks and carry out significant assaults, like Distributed Denial of Service (DDoS) attacks on crucial infrastructures.

In the current interconnected environment of personal and business systems, it is common for individuals to bring their devices to work, thereby increasing the risk of individual malware infections spreading to more extensive corporate networks. Consequently, it is crucial for organizations to have a clear understanding of the characteristics of malicious software and to educate users on the various types of malware protection they can incorporate into their security protocols to protect their physical and digital assets. There are several ways that a system can be infected by malware, but common ways are:
 \begin{itemize}
 	\item Opening malicious email attachments
 	\item Installing untrustworthy applications without antivirus software poses significant security risks 
 	\item Visiting sites infected with malware
 	\item Downloading PDF, video, malicious music files
 	\item Not patching/updating the operating system and applications
 \end{itemize}

Consequently, malware prevention depends on two critical factors: user awareness of malicious threats and system properties. Exploits commonly target vulnerabilities in widely used software, including web browsers, Java, Adobe Flash Player, and Microsoft Office, to compromise devices. Regular software updates address these vulnerabilities by applying patches, rendering them inaccessible to potential exploits. Accordingly, software version, operating system platform sub-release, firewall status, and other system properties and configurations play an essential role in detecting whether a computer system is vulnerable to malware.

Many organizations invest severely in their cyber-security programs, only to discover they overlooked or lacked something significant when a cyber-attack occurs. For this reason, it’s imperative to regularly test security systems, processes, and personnel to identify vulnerabilities and gaps before somebody with wicked intentions finds them first. As a result, detecting a vulnerable system can play a critical role in helping organizations and individuals better understand the vulnerabilities of their systems and put the right tools and processes in place to alleviate them before it’s too late.

Our goal in this study is to forecast the susceptibility of Microsoft computers to malware by analyzing a range of machine characteristics. We used the "Microsoft Malware Prediction" dataset available on Kaggle to accomplish this. This dataset contains telemetry data outlining these characteristics and the corresponding instances of machine infections obtained from reports produced by Microsoft's endpoint protection solution, Windows Defender, from 8,921,483 Microsoft Windows machines. The sampling method used to compile this dataset conforms to specific business restrictions related to user privacy and the operational duration of the machines. While identifying malware is inherently a matter of time series, it is further complicated by factors such as the addition of new machines, the online and offline status of machines, the application of patches, and the installation of new operating systems. Additionally, this dataset does not accurately represent the machines of Microsoft customers; it has been sampled to include a disproportionately higher number of machines infected with malware. Here are the specific objectives:

\begin{enumerate}
    \item How can ML help to predict which machines are vulnerable to malware and thereby protect them and prevent it from further damage before it happens?
    \item Assess the likelihood of a Windows machine becoming infected by various malware families by analyzing its distinct properties.
    \item Do states of machines play a role in getting infected by malware?
    \item Which states have a critical impact on preventing malware attacks?
    \item Evaluate and compare the effectiveness of various classification methods in detecting malware.
\end{enumerate}

\section{Methods}\label{methods}
\subsection{Data description and pre-processing}
In this paper, the ``Microsoft Malware Prediction" dataset from Kaggle is used. This dataset contains various attributes and the infection status of 8,921,483 machines. Each entry in the dataset represents a Microsoft system, uniquely identified by the \colorbox{lightgray}{MachineIdentifier}. The \colorbox{lightgray}{HasDetections} field serves as an indicator of whether a particular machine has been infected by malware. For each machine, 81 distinct features are provided. Additionally, columns with missing or self-explanatory names are labeled as ``NA." The problem is framed as a two-class classification task. Before model training, we provide an overview of the dataset.


\subsubsection{Descriptive Analysis}
There are approximately 9 million records. A basic analysis indicates that the dataset has noticeable amounts of records with missing values. First, we investigate the percentage of missing values in each feature to deal with missing values in the data set. The features, \colorbox{lightgray}{DefaultBrowsersIdentifier}, \colorbox{lightgray}{PuaMode}, and \colorbox{lightgray}{Census\_ProcessorClass} have the highest missing values (more than $\%90$). Furthermore, as the features with the second highest missing values, the following features contain $\%60$ to $\%90$ of missing values.
\\\colorbox{lightgray}{Census\_InternalBatteryType},
\colorbox{lightgray}{Census\_IsFlightingInternal},
\colorbox{lightgray}{Census\_PThresholdOptIn},\\ \colorbox{lightgray}{Census\_IsWIMBootEnabled}

In addition, we examine the count of the values of each feature (skewness). The values in the following 26 features are skewed toward one category (more than $\%90$ of the values).
\\
\colorbox{lightgray}{ProductName}, \colorbox{lightgray}{IsBeta}, \colorbox{lightgray}{RtpStateBitfield}, \colorbox{lightgray}{IsSxsPassiveMode}, \colorbox{lightgray}{DefaultBrowsersIdentifier}, \colorbox{lightgray}{AVProductsEnabled}, \colorbox{lightgray}{HasTpm}, \colorbox{lightgray}{Platform}, \colorbox{lightgray}{Processor}, \colorbox{lightgray}{OsVer}, \colorbox{lightgray}{IsProtected}, \colorbox{lightgray}{AutoSampleOptIn}, \colorbox{lightgray}{PuaMode}, \colorbox{lightgray}{SMode}, \colorbox{lightgray}{Firewall}, \colorbox{lightgray}{UacLuaenable}, \colorbox{lightgray}{Census\_DeviceFamily}, \colorbox{lightgray}{Census\_ProcessorClass}, \colorbox{lightgray}{Census\_HasOpticalDiskDrive}, \colorbox{lightgray}{Census\_OSArchitecture}, \colorbox{lightgray}{Census\_IsPortableOperatingSystem}, \colorbox{lightgray}{Census\_IsFlightsDisabled}, \colorbox{lightgray}{Census\_FlightRing}, \colorbox{lightgray}{Census\_IsVirtualDevice}, \colorbox{lightgray}{Census\_IsPenCapable}, \colorbox{lightgray}{Census\_IsAlwaysOnAlwaysConnectedCapable}.

As the first step of our data clarification, we removed the abovementioned features and reduced the number of features to 51 variables. Further investigation of the current dataset shows that many missing values still exist. Since we have computational resource limitations, we decided to drop instances with missing values. Hence, the obtained dataset contains 51 features of $3501407$ machines. This dataset is balanced since $\%51.2$ of machines belong to class $1$, and $\%48.8$ belongs to class $0$.

The data set has three features: categorical, numerical, and binary-valued. Figure 1 demonstrates the distribution of features in three classes.

\begin{figure}[H]
	\centering
	\fbox{\includegraphics[width=0.8\textwidth]{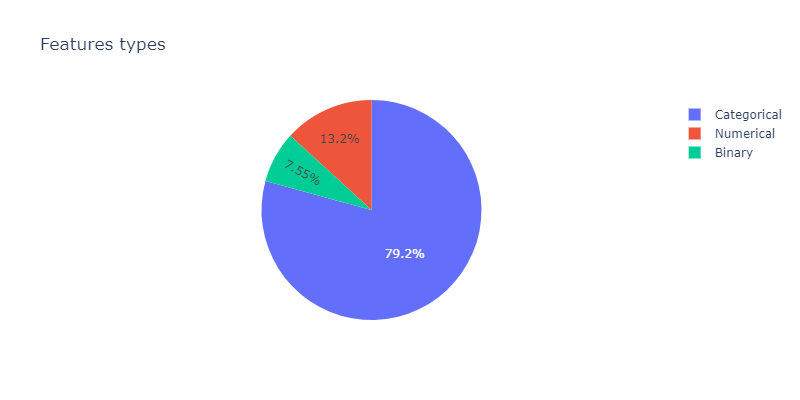}}
	\caption{distribution of features in three classes}
\end{figure}

Furthermore, as Figure 1 shows, most of the features are categorical, and we need to obtain a method to treat them depending on the cardinality of each feature's categories. Figure 2 shows the top $10$ features with high categorical values.

\begin{figure}[H]
	\centering
	\fbox{\includegraphics[width=0.8\textwidth]{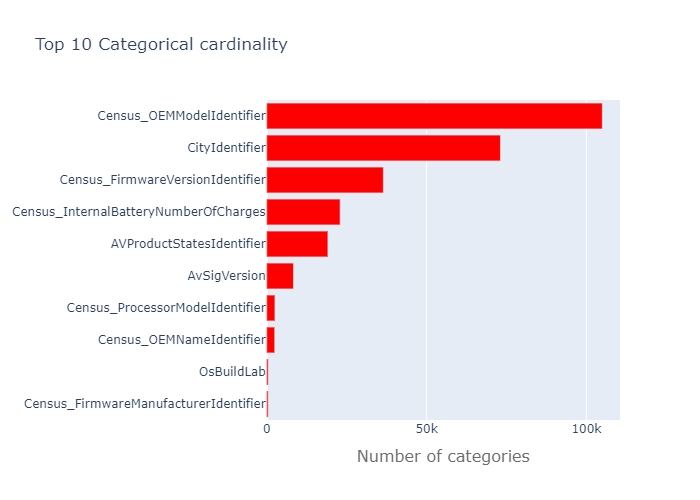}}
	\caption{Top $10$ features that have high categorical cardinality}
\end{figure}

We begin our analysis with features with high categorical cardinality.
\subsubsection{Feature engineering}

Considering the obtained information from the dataset, feature engineering is challenging. To this aim, the binary-valued features remain intact. Our strategy is to engineer the remaining features to be scaled to lower values. In other words, the numerical-valued features are normalized via the Min-Max method; Categorical features with lower than $5$ categories are encoded by the One-Hot method; Categorical features with categories between $5$ and $20$ are encoded through factorizing; and categorical features with more than $20$ categories are encoded via Frequency encoding method because its not feasible to use LabelEncoding technique when cardinalities of categories in categorical features are very high (to avoid having values closed to zero we apply Min-Max normalization after frequency encoding).

Although feature engineering was challenging, there were some interesting categorical features that we tried to treat nicely. One of the interesting findings is related to the features \colorbox{lightgray}{CountryIdentifier} and \colorbox{lightgray}{CityIdentifier}. The dataset contains the records of $215$ countries and $73121$ cities. It is reasonable to consider that each country has unique city IDs since a city can't belong to more than one country. The cross-match classification of country and city IDs denoted $75921$ city IDs! Therefore, there exist $2109$ cities that belong to more than one country. In conclusion, we found \colorbox{lightgray}{CityIdentifier} feature misleading and dropped it from the dataset.

The features \colorbox{lightgray}{AvSigVersion}, \colorbox{lightgray}{EngineVersion}, \colorbox{lightgray}{AppVersion}, \colorbox{lightgray}{OsBuildLab}, and \colorbox{lightgray}{Census\_}\\\colorbox{lightgray}{OSVersion} have segmented categories in \#.\#.\#.\# or \#.\#.\#.\#.\# format. We found that analyzing these features by each segment in their category reduces the feature's categorical cardinality and gives us more opportunity to encode them appropriately. We deal with each of the abovementioned features as the following:

\begin{itemize}
    \item \colorbox{lightgray}{AvSigVersion}: The first and last segments are the same for all records, so we ignore them. The second segment has $51$ categories, and most values belong to two categories; we combine the remaining as the third category and encode it via the one-hot method. The third segment has $2725$ categories we encode using the frequency method.
    \item \colorbox{lightgray}{EngineVersion}: The first and second segments are the same for all records, so we ignore them. The third segment has $60$ categories, and most values belong to two categories; we combine the remaining as the third category and encode it via the one-hot method. The fourth segment has seven categories, and most values belong to one category; we combine the remaining as the second category and encode it via the one-hot method.
    \item \colorbox{lightgray}{AppVersion}: The first segment is the same for all records; so, we ignore it. The second, third, and fourth segments have $15$, $35$, and $78$ categories, respectively, and most of the values belong to one category; for each segment, we combine the remaining as the second category and encode it via the one-hot method.
    \item \colorbox{lightgray}{OsBuildLab}: The first, second, and fifth segments have $54$, $222$, and $226$ categories that most of the values belong to seven, four, and four categories, respectively; for each segment, we combine the remaining as another category and encode it via the factorization method. The third segment has three categories, and most of the values belong to two categories; we combine the remaining into the second category and encode it via the one-hot method. The fourth segment is the same as \colorbox{lightgray}{Census\_OSBranch}; so, we ignore it.
    \item \colorbox{lightgray}{Census\_OSVersion}: The first and second segments are the same for all records. The third segment is the same as \colorbox{lightgray}{Census\_OSBuildNumber}. The fourth segment is the same as \colorbox{lightgray}{Census\_OSBuildRevision}. Therefore, we drop this feature.
\end{itemize}

Combining similar categories into one category or combining less frequent categories as a separate category is applied to the remaining features whenever their application is reasonable. After pre-processing the data set and engineering its features, we obtained $97$ encoded features.

\subsection{Models for Malware Detection}

Various ML methods in the literature are employed for malware detection problems primarily based on ensemble methods. Since one of our objectives is to compare different methods' performance on this problem, we develop some single classifiers - Gaussian NB, Logistic regression, DT - and ensemble methods - RF, Gradient Boosting (XGBoost, LightGBM), and Stacking. This section describes each technique and its tuned parameters and accuracy. We apply the train-test split for each model with a $0.7:0.3$ portion. The stand-alone classifiers are chosen due to their lower computational time (i.e., we have a large data set).

\subsubsection{Model 1: Bayesian Classifier}

NB methods are a group of supervised learning algorithms that rely on Bayes' theorem, incorporating the ``naive" assumption that all features are conditionally independent of one another, given the class variable's value. Assuming the features likelihood is Gaussian, we implement the Gaussian NB classifier. The training and testing accuracy are $\%57.54$ and $\%57.52$, respectively. The results show this model is slightly better than random guessing. Hence, its performance is insufficient for this problem as a stand-alone model.

\subsubsection{Model 2: Logistic Regression}

Logistic regression is a machine learning method that employs the logistic function to predict a binary dependent variable in its simplest form, though various more advanced extensions of the model are available. As the first trial, the model is built by setting the stochastic average gradient (sag) as the solver and a maximum iteration of 1000. Then, applying 5-fold cross-validation considering maximum iteration as $\left[100, 500, 1000, 1500, 2000, 2500, 3000\right]$, the inverse of regularization strength (C) as $\left[10^{-4}, 10^{-3}, 10^{-2}, 10^{-1}, 1, 10, 10^2, 10^3, 10^4\right]$, and Solver: $\left["sag", "saga"\right]$\footnote{The saga solver is a modification of the sag solver that additionally supports L1 regularization. As a result, it is the preferred solver for sparse multinomial logistic regression and is well-suited for handling very large datasets.} The best model is trained by "saga" as a solver, $C=1$, and a maximum iteration of $500$. The results showed $\%61.79$ and $\%61.83$ training and testing accuracy, respectively. Although we observed a slight improvement in accuracy, the performance was inadequate for this critical problem.

\subsubsection{Model 3: Decision Tree}
The DT was the third model selected for training. As a predictive model, a DT progresses from observations about a particular item (depicted by the branches) to predictions about the target value of that item (represented by the leaves). In the first training trial, a DT overfitted the training set ($\%99.99$ training accuracy and $\%55.58$ test accuracy). Therefore, we apply a grid-search method by 3-fold cross-validation considering both criteria (gini-index and entropy), maximum tree depth between $8$ and $25$, kind of split as "best" and "random," and maximum features number as $\left["sqrt," "log2", None\right]$. The best model is trained by, "entropy" as criteria, $12$ as maximum depth of tree, $"random"$ as splitter, and maximum features number of $"None"$, with best score $0.6224$. The results showed $\%62.75$ and $\%62.35$ training and testing accuracy, respectively. DTs showed slightly higher accuracy than logistic regression.

\subsubsection{Model 4: Stacking}

Since each classifier performed like a weak classifier, we tried building ensemble methods. The first idea was to build a stacking classifier with each of the previously built models. Stacked generalization involves combining the outputs of multiple individual estimators and using a final classifier to make the overall prediction. By stacking, the strengths of each estimator are leveraged, as their outputs are used as inputs for the final model to produce the final prediction. Therefore, we constructed each of the optimal individual models -- ``Decision Tree", ``Logistic Regression", ``Gaussian Naive Bayes", ``Extra-trees Classifier"\footnote{In highly randomized trees, randomness affects both the selection of features and the way splits are made. Unlike random forests, where the most discriminative thresholds are chosen, highly randomized trees generate random thresholds for each feature and select the best one as the splitting rule. This method helps further reduce model variance, but it introduces a small increase in bias.}, and ``Random Forest" -- using their respective tuned parameters. The XGBoost is considered the final classifier (decision tree and logistic regression had insufficient performance, and SVM had more than a day of training time). Training this model showed $\%63.46$ and $\%62.89$ training and testing accuracy, respectively. Training of this model takes more than 8 hours. Therefore, we did not conduct further analysis on this model.

\subsubsection{Model 5: Gradient Boosting - XGBoost}

As an alternative approach, tree-based ensemble methods were selected, as decision trees outperformed other models. XGBoost, short for eXtreme Gradient Boosting, is a gradient boosting framework designed to be scalable, portable, and distributed, specifically for gradient boosting methods (GBM, GBRT, GBDT). It utilizes gradient-boosted decision trees that are specifically optimized for both performance and speed. In gradient boosting, new models are trained to predict the residuals or errors of previous models, and these predictions are combined to make the final output. This approach is known as gradient boosting because it uses the gradient descent algorithm to minimize loss as new models are added. However, XGBoost does not natively handle categorical features, accepting only numerical inputs like Random Forests, so categorical data must be encoded before being used with XGBoost.

The model was constructed by dividing the training data into a training set and a validation set with an $0.8:0.2$ split ratio. The first trial of training XGBoost has overfitted the training set ($\%99.99$ training accuracy and $\%61.08$ test accuracy) with training 10 minutes of training time. Although this model is overfitted, training time was promising. Considering this observation, we tuned the parameters of XGBoost. The tuned model is built according to the settings in Table 1:

\begin{table}[H]
\centering
\caption{Parameter setting of XGBoost}
\begin{tabular}{|cc|cc|} 
\hline
\textbf{Parameter} & \textbf{Value} & \textbf{Parameter} & \textbf{Value}       \\ 
\hline
n\_estimators       & $1000$          & booster             & $"gbtree"$           \\
eta                 & $0.1$          & gamma               & $1$                  \\
max\_depth          & $6$            & sampling\_method    & $"gradient\_based"$   \\
reg\_lambda         & $0.15$         & reg\_alpha          & $0.15$               \\
max\_bin        & $1024$       & objective           & $"binary:logistic"$  \\
\hline
\end{tabular}
\end{table}

We built a model concerning the importance of the feature that noticeable change in the performance was not observed. Considering the evaluation set as a training-validation set and evaluation metrics as $"error"$ and $"log loss"$, we monitor the training loss and validation loss. Our observation denoted that after almost $400$ epoch, the model tends to overfit. Accordingly, by setting an early stopping, we achieve $\%66.08$ and $\%64.52$ accuracy for training and testing, respectively. 

\subsubsection{Model 6: Gradient Boosting - LightGBM}

LightGBM employs a unique technique called ``Gradient-based One-Sided Sampling (GOSS)" to filter data instances when determining the threshold value. In contrast, ``XGBoost" uses a pre-sorted algorithm and a histogram-based approach to determine the optimal split.`` LightGBM" can natively process categorical features by accepting feature names directly, avoiding the need for one-hot encoding, which significantly improves speed. It uses a specialized algorithm to identify the split value for categorical features. Key advantages of LightGBM include enhanced training speed, higher efficiency, reduced memory utilization, refined  accuracy, ability to leverage parallel and GPU computing, and capability to process large datasets.

To develop this model, we partitioned the training dataset into training and validation subsets using an $0.8:0.2$ split. Setting the objective and boosting of LightGBM as "binary" and "Gradient Boosting Decision Tree (gbdt)," the initial model was built. The observation was promising since it had a lower training time (5 minutes) and performance similar to XGBoost. Considering this observation, we tuned the related parameters of LightGBM. Table~\ref{tab2} denotes the tuned parameters.

\begin{table}[H]
\centering
\caption{LightGB parameter setting}\label{tab2}
\begin{tabular}{|cc|cc|} 
\hline
\textbf{Parameters} & \textbf{Value} & \textbf{Parameters} & \textbf{Value}  \\ 
\hline
n\_estimators       & $1000$         & boosting            & $"gbdt"$        \\
learning\_rate      & $0.05$         & objective           & $"binary"$      \\
max\_depth          & $10$           & feature\_fraction   & $0.9$           \\
lambda\_l1          & $0.15$         & lambda\_l2          & $0.15$          \\
num\_leaves         & $2048$         & bagging\_freq       & $8$             \\
bagging\_fraction   & $0.8$          & bagging\_seed       & $15$            \\
\hline
\end{tabular}
\end{table}

We built a model concerning the importance of the feature that noticeable change in the performance was not observed. Considering the evaluation set as a training-validation set and evaluation metrics as $"error"$ and $"log loss"$, we monitor the training loss and validation loss. Our observation denoted that after almost $200$ epoch, the model tends to overfit. Accordingly, by setting an early stopping, we achieve $\%69.68$ and $\%64.52$ accuracy for training and testing, respectively.

\subsubsection{Other methods}

Other than the above mentioned models, we tried to train models that we decided not to proceed with due to high training time or other restrictions. 

Support vector machines (SVMs) are among the most reliable prediction methods, rooted in statistical learning theory. Given a set of training examples classified into one of two categories, the SVM algorithm constructs a model that classifies new data points into one of the categories, functioning as a non-probabilistic binary linear classifier. SVMs work by projecting training data in a high-dimensional feature space and maximizing the boundary between the two classes. We experimented with SVMs using different kernels (linear, RBF, sigmoid, and polynomial). However, owing to the substantial amount of data, the training process for SVMs was time-consuming, taking several days without yielding an initial model. As a result, we decided not to continue with SVM.

Random forests are an ensemble learning method utilized for tasks such as classification, regression, and other predictive modeling applications. The method functions by creating multiple decision trees during training and making predictions based on the majority class for classification or the average prediction for regression from the individual trees. First trial of training a decision tree is overfitted the training set ($\%99.99$ training accuracy and $\%62.78$ test accuracy) with 20 minutes training time. Therefore, we decided to build a forest of the best trees we built as Model 3 with 100 estimators. The training results denoted $\%63.67$ and $\%62.91$ training and testing accuracy with 2 hours of training time. This initial observation indicates that tree-based ensemble methods perform better than individual methods. However, the long training time of random forests was a big restriction for us when proceeding with this model.

This section explained the ML tools we used to fit our dataset and their tuned parameters. In the next section, we investigate the obtained results deeply and compare them with each other.

\section{Computational Results}\label{results}
From the discussion in the Introduction, the critical task in this problem is to recognize malware as a safe program. This is equivalent to \textit{Miss} or \textit{type II error} of the classification method. This does not mean detecting a safe program as malware (\textit{Type II error}) is not essential. However, in the malware detection problem, the \textit{type II error} is more significant than \textit{Type II error}. With this objective in mind, we compare the results of each model we developed to detect malware before an infected machine.

Table 3 denotes the classification result of each classifier. Since the dataset is balanced, no significant difference in precision and recall is observed for each model. With these results in mind, it can be inferred from Table 1 that, in all classifiers, the proportion of malware identifications identified by the machine is the same as the proportion of actually identified malware on machines determined by the model correctly. However, their percentage varies for each model, which denotes the strength of each model.

\begin{table}[H]
\centering
\caption{Classification report of each model}
\resizebox{\columnwidth}{!}{%
\begin{tabular}{|c|cc|cc|cc|} 
\hline
\multirow{2}{*}{\textbf{Classifier }}   & \multicolumn{2}{c|}{\textbf{Precision (avg)}}  & \multicolumn{2}{c|}{\textbf{Recall (avg)}}                            & \multicolumn{2}{c|}{\textbf{ F\_1 score (avg)}}\\ 
\cline{2-7}
& \multicolumn{1}{c|}{\textbf{Macro}}        & \textbf{Weighted}     & \multicolumn{1}{c|}{\textbf{Macro}}        & \textbf{Weighted}     & \multicolumn{1}{c|}{\textbf{Macro}}        & \textbf{Weighted}     \\ 
\hline
Gaussian Naive Bayes & $0.60$  & $0.60$ & $0.57$ & $0.58$ & $0.54$ & $0.54$\\
\textcolor[rgb]{0.129,0.129,0.129}{Decision Tree} & \textcolor[rgb]{0.129,0.129,0.129}{$0.62$} & \textcolor[rgb]{0.129,0.129,0.129}{$0.62$} & \textcolor[rgb]{0.129,0.129,0.129}{$0.62$} & \textcolor[rgb]{0.129,0.129,0.129}{$0.62$} & \textcolor[rgb]{0.129,0.129,0.129}{$0.62$} & \textcolor[rgb]{0.129,0.129,0.129}{$0.62$}  \\
Logistic Regression                               & \textcolor[rgb]{0.129,0.129,0.129}{$0.62$} & \textcolor[rgb]{0.129,0.129,0.129}{$0.62$} & \textcolor[rgb]{0.129,0.129,0.129}{$0.62$} & \textcolor[rgb]{0.129,0.129,0.129}{$0.62$} & \textcolor[rgb]{0.129,0.129,0.129}{$0.62$} & \textcolor[rgb]{0.129,0.129,0.129}{$0.62$}  \\
Stacking                                          & \textcolor[rgb]{0.129,0.129,0.129}{$0.63$} & \textcolor[rgb]{0.129,0.129,0.129}{$0.63$} & \textcolor[rgb]{0.129,0.129,0.129}{$0.63$} & \textcolor[rgb]{0.129,0.129,0.129}{$0.63$} & \textcolor[rgb]{0.129,0.129,0.129}{$0.63$} & \textcolor[rgb]{0.129,0.129,0.129}{$0.63$}  \\
XGBoost                                           & $0.64$                                     & $0.65$                                     & $0.64$                                     & $0.65$                                     & $0.64$                                     & $0.65$                                      \\
LightGBM                                          & $0.64$                                     & $0.65$                                     & $0.64$                                     & $0.65$                                     & $0.64$                                     & $0.65$                                      \\
\hline
\end{tabular}
}
\end{table}

Table 4 denotes each model's confusion matrix and corresponding training and testing accuracy. The confusion matrix is generated by utilizing the model's predictions on the test set, showing the size and distribution of outcomes across the various classes. Investigation of the accuracy denotes that ensemble methods, especially gradient-based frameworks, have higher training and testing accuracy than each model.

\begin{table}[H]
\centering
\caption{Confusion matrix and accuracy metric of each model}
\resizebox{\columnwidth}{!}{%
\begin{tabular}{|c|cccc|c|c|} 
\hline
\textbf{Classifier}& \textbf{TP} & \textbf{FP}& \textbf{TN} & \textbf{FN}& \textbf{Training Accuracy} & \textbf{Test Accuracy}  \\ 
\hline
Gaussian Naive Bayes & \textcolor[rgb]{0.129,0.129,0.129}{$466329$} & \textcolor[rgb]{0.129,0.129,0.129}{$371352$} & \textcolor[rgb]{0.129,0.129,0.129}{$160362$} & \textcolor[rgb]{0.129,0.129,0.129}{$91392$}  & $0.58$& $0.58$ \\
\textcolor[rgb]{0.129,0.129,0.129}{Decision Tree} & \textcolor[rgb]{0.129,0.129,0.129}{$353146$} & \textcolor[rgb]{0.129,0.129,0.129}{$205692$} & \textcolor[rgb]{0.129,0.129,0.129}{$326022$} & \textcolor[rgb]{0.129,0.129,0.129}{$204575$} & $0.63$& $0.62$ \\
Logistic Regression & \textcolor[rgb]{0.129,0.129,0.129}{$368032$} & \textcolor[rgb]{0.129,0.129,0.129}{$226143$} & \textcolor[rgb]{0.129,0.129,0.129}{$305571$} & \textcolor[rgb]{0.129,0.129,0.129}{$189689$} & $0.62$ & $0.62$                  \\
Stacking & \textcolor[rgb]{0.129,0.129,0.129}{$354788$} & \textcolor[rgb]{0.129,0.129,0.129}{$201375$} & \textcolor[rgb]{0.129,0.129,0.129}{$330339$} & \textcolor[rgb]{0.129,0.129,0.129}{$202933$} & $0.63$                     & $0.63$                  \\
XGBoost& $366163$& $194961$ & $336753$& $191558$ & $0.66$& $0.64$                  \\
LightGBM& $365576$& $194385$  & $337329$& $192145$& $0.69$& $0.65$\\ 
\hline
\textbf{Training set}& \multicolumn{6}{c|}{2,033,612}\\
\textbf{Test set} & \multicolumn{6}{c|}{1,089,435}\\
\textbf{Validation set} & \multicolumn{6}{c|}{508,403}\\
\hline
\end{tabular}
}
\end{table}

An interesting observation in Table 4 is related to the Gaussian Naive Bayes (GNB) model results. The results denote that GNB could detect most malware attacks, given the state of each machine. However, GNB predicted most safe programs as malware; accordingly, the number of malware cases identified as safe programs is lower. This observation denotes that GNB behaved more liberally than the other classifiers. Although GNB addresses the critical issue in malware detection, it labels most safe programs as malware, which might cause users dissatisfaction.

The other individual classifiers are more conservative than GNB. Although DT and Logistic Regression (LR) behavior has increased their accuracy, they cannot deal with the major issue of this problem (i.e., lower FN).

Among the ensemble models, XGBoost and LightGBM perform better than stacking. One reason for that might be the different attitudes of each weak classifier. LightGBM's higher accuracy is due to its ability to detect safe programs.

\begin{figure}[H]
	\centering
\includegraphics[width=0.9\textwidth]{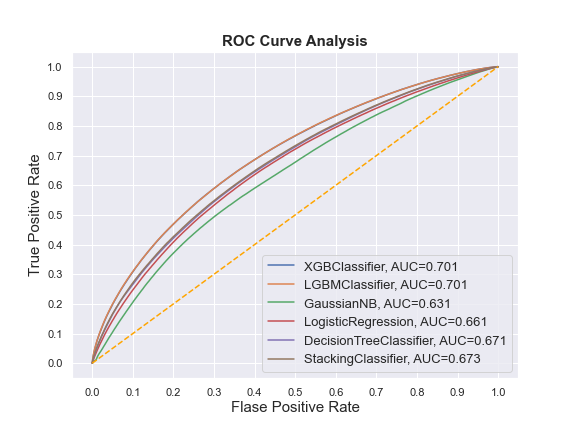}
	\caption{ROC curve of the methods}
\end{figure}

Similar results are illustrated in Figure 3, which presents the ROC curve performance analysis. Observing the figure shows that GNB performs less than the other classifiers. DT, LR, and stacking methods perform similarly and are better than GNB. XGBoost and LightGBM perform better than the others. This figure shows that the ensemble methods outperform each method for malware detection problems and are more conservative than the other models.

This figure also illustrates the similar trends in the variation of specificity, sensitivity (recall), and precision. When the actual ROC curve follows a smooth function, the accuracy of statistical inferences drawn from the empirical ROC curve is lower compared to those made using a model-based estimator, assuming the model is correctly specified.

Further analysis of the applied ensemble methods focuses on the importance of features in predicting each model's outcome. The results show that the \colorbox{lightgray}{AVProductStatesIdentifier} serves as a major factor in the XGBoost predictions. Additionally, \colorbox{lightgray}{AppVersion} and \colorbox{lightgray}{AvSigVersion} are also found to be important for predicting the target variable. In the LightGBM model, \colorbox{lightgray}{AvSigVersion} is highly effective for predicting the target variable. Moreover, \colorbox{lightgray}{Census\_FirmwareVersionIdentifier}, \colorbox{lightgray}{CountryIdentifier}, and \colorbox{lightgray}{Census\_SystemVolumeTotalCapacity}, have a significant impact on predicting the target variable in this model.

\subsection{Discussion on results}
Malware attacks have become an increasing concern for data security and integrity for individual users and organizations utilizing Windows systems. With computers' increasing role in our daily lives, threats, and malicious software are designed to sabotage critical systems. ML tools have been utilized as promising methods to prevent malware from infecting machines. Considering Microsoft's provided dataset for malware detection, we tried to build an ML model to detect malware based on a machine's state.

The advancement of the ML community has provided powerful classification algorithms that can be utilized to predict malware before a machine is infected. We developed a Gaussian Naive Bayes model, a statistical supervised learning method, to predict the probability of infection based on the status of each machine. Although this model is slightly better than random guessing, it acted as a conservative classifier. As we present the results of each model, this model tends to predict that most of the machines are vulnerable to getting infected via malware. Since the most critical task of a robust model in malware detection is to have a lower type II error (i.e., label an unsafe machine as a protected machine), this behavior of GNB seems beneficial due to its lower false-positive rate. However, this model leads to the Microsoft users' discontent. Detecting most machines as vulnerable systems to malware makes the system inflexible and restricts the users' jobs on machines.

A logistic regression model has been built to resolve GNB's shortcomings. After tuning the parameters and cross-validation, the best logistic regression model is identified. The results demonstrate an improvement in accuracy; however, the type II error rate has increased. This suggests that logistic regression is more lenient compared to GNB. While GNB shows a higher type I error and a lower type II error, logistic regression results in a lower rate of type I errors while having a higher rate of type II errors in comparison. These findings highlight the limitations of linear classifiers on the given dataset.

Observing the low efficiency of the logistic regression, as a linear classifier, our next choice was the decision tree. The best-fitted decision tree classifier is obtained after fine-tuning its parameters and the cross-validation. The decision tree’s performance is comparable to that of logistic regression. However, it results in lower type I and higher type II errors compared to logistic regression. Our analysis indicates that the decision tree balances types I and II errors. This suggests that tree-based methods are a good fit for the dataset; however, the performance is limited due to the inefficiency of using a single tree for classification.

The performance of the single classifiers denotes they perform well only in a part of the dataset. Therefore, relying on one of them as the final model is unreasonable. Hence, we tried to build ensemble models. As the first approach, we built a stacking model of the developed best single classifiers. Considering GNB, logistic regression, decision tree, extra-tree, and random forests as the weak classifiers and XGBoost as the meta-learner, the stacking model is built. Although this model's type I and type II errors are similar to the single decision tree classifier, their accuracy is higher than the single decision tree. Although the training time of this model was very high, its result is a compelling reason to try the most influential ensemble methods based on gradient boosting. Before developing these classifiers, we built a random forests classifier with a higher training time, the same as the stacking method. Hence,  we were not able to conduct further analysis on them.

To resolve the drawback of the previous models, we developed two gradient boosting-based models, XGBoost and LightGBM. These methods are much faster and memory-efficient than the stacking model. LightGBM employs an innovative approach called GOSS to refine data instances for finding split values, whereas XGBoost relies on pre-sorted and histogram-based algorithms to find the optimal split. LightGBM outperforms XGBoost in terms of speed and memory efficiency. The results indicate that LightGBM achieves lower type I and type II errors and higher accuracy. Furthermore, comparing ROC curves among the classifiers reveals that LightGBM has a higher ``Area Under the Curve (AUC)" and surpasses the other models' overall performance.

The result of gradient-based methods denotes that \colorbox{lightgray}{AVProductStatesIdentifier}, \colorbox{lightgray}{AppVersion}, \colorbox{lightgray}{Census\_SystemVolumeTotalCapacity}, \colorbox{lightgray}{AvSigVersion}, \colorbox{lightgray}{Census\_FirmwareVersionIdentifier}, and \colorbox{lightgray}{CountryIdentifier} are the features that play significant role in predicting the target variable.

\section{Conclusion}\label{conclusion}
Malware is currently one of the most severe security threats on the Internet. Many common online issues, such as ``spam emails" and ``denial-of-service" attacks, often stem from malware. Compromised computers are frequently organized into botnets -- networks of infected machines controlled by attackers -- and these networks are used to launch various malicious attacks. To address the constant evolution of malware, new techniques are required to detect and mitigate the damage they cause. In this paper, we developed five machine learning models, including ``Gaussian Naive Bayes", ``Logistic Regression", and ``Decision Trees" as individual models, gradient-based ensemble models (XGBoost and LightGBM), and a stacking model, which combines Gaussian Naive Bayes, logistic regression, decision trees, extra-trees, and random forests as weak learners, with XGBoost serving as the meta-learner.

The results indicate that LightGBM surpasses the other models in terms of overall performance, with faster training speeds and greater memory efficiency. However, despite achieving the highest accuracy, the model's practical effectiveness is limited, as it only reaches 65\% accuracy. This suggests that the model is not sufficiently complex to effectively detect malware. We believe that this may be due to the exclusion of all missing values, which was necessary because of computational resource constraints. With more robust computing systems, including the missing values could potentially lead to higher accuracy. Additionally, generating new features from the existing dataset or acquiring data from different machines could increase model complexity. Therefore, further feature engineering may enhance the model's performance. Moreover, exploring other hyperparameters in the ensemble models could yield models with improved accuracy.


\section*{Acknowledgment}
Not applicable.

\section*{Conflict of interest} 
The authors have no conflicts of interest to disclose.

\section*{Funding} 
The authors received no financial support for this paper's research, authorship, and publication.

\section*{Data Availability Statement}
The dataset can be downloaded via the following link: \hyperlink{https://www.kaggle.com/c/microsoft-malware-prediction}{https://www.kaggle.com/c/microsoft-malware-prediction}

\bibliographystyle{unsrt}
\bibliography{bib}

\begin{thebibliography}{10}

\bibitem{Aftabi16012025}
Navid Aftabi, Dan Li, and Thomas~C. Sharkey.
\newblock An integrated cyber-physical framework for worst-case attacks in industrial control systems.
\newblock {\em IISE Transactions}, 0(0):1--19, 2025.

\bibitem{tahir2018study}
Rabia Tahir.
\newblock A study on malware and malware detection techniques.
\newblock {\em International Journal of Education and Management Engineering}, 8(2):20, 2018.

\bibitem{aftabi2025sd}
Navid Aftabi, Nima Moradi, Fatemeh Mahroo, and Farhad Kianfar.
\newblock Sd-abm-ism: An integrated system dynamics and agent-based modeling framework for information security management in complex information systems with multi-actor threat dynamics.
\newblock {\em Expert Systems with Applications}, 263:125681, 2025.

\bibitem{al2021efficient}
Neamat Al~Sarah, Fahmida~Yasmin Rifat, Md~Shohrab Hossain, and Husnu~S Narman.
\newblock An efficient android malware prediction using ensemble machine learning algorithms.
\newblock {\em Procedia Computer Science}, 191:184--191, 2021.

\bibitem{kilgallon2017improving}
Sean Kilgallon, Leonardo De~La~Rosa, and John Cavazos.
\newblock Improving the effectiveness and efficiency of dynamic malware analysis with machine learning.
\newblock In {\em 2017 Resilience Week (RWS)}, pages 30--36. IEEE, 2017.

\bibitem{ye2017survey}
Yanfang Ye, Tao Li, Donald Adjeroh, and S~Sitharama Iyengar.
\newblock A survey on malware detection using data mining techniques.
\newblock {\em ACM Computing Surveys (CSUR)}, 50(3):1--40, 2017.

\bibitem{aslan2020comprehensive}
{\"O}mer~Aslan Aslan and Refik Samet.
\newblock A comprehensive review on malware detection approaches.
\newblock {\em IEEE access}, 8:6249--6271, 2020.

\bibitem{aboaoja2022malware}
Faitouri~A Aboaoja, Anazida Zainal, Fuad~A Ghaleb, Bander Ali~Saleh Al-Rimy, Taiseer Abdalla~Elfadil Eisa, and Asma Abbas~Hassan Elnour.
\newblock Malware detection issues, challenges, and future directions: A survey.
\newblock {\em Applied Sciences}, 12(17):8482, 2022.

\bibitem{gopinath2023comprehensive}
Mohana Gopinath and Sibi~Chakkaravarthy Sethuraman.
\newblock A comprehensive survey on deep learning based malware detection techniques.
\newblock {\em Computer Science Review}, 47:100529, 2023.

\bibitem{goyal2020pipeline}
Manish Goyal and Raman Kumar.
\newblock The pipeline process of signature-based and behavior-based malware detection.
\newblock In {\em 2020 IEEE 5th International Conference on Computing Communication and Automation (ICCCA)}, pages 497--502. IEEE, 2020.

\bibitem{galal2016behavior}
Hisham~Shehata Galal, Yousef~Bassyouni Mahdy, and Mohammed~Ali Atiea.
\newblock Behavior-based features model for malware detection.
\newblock {\em Journal of Computer Virology and Hacking Techniques}, 12:59--67, 2016.

\bibitem{niraj2022performance}
Shashi~Prakash Niraj and Ajit~Kumar Tiwari.
\newblock Performance analysis of signature based and behavior based malware detection.
\newblock {\em Research Journal of Engineering Technology and Medical Sciences (ISSN: 2582-6212)}, 5(04), 2022.

\bibitem{bahador2019hlmd}
Mohammad~Bagher Bahador, Mahdi Abadi, and Asghar Tajoddin.
\newblock Hlmd: a signature-based approach to hardware-level behavioral malware detection and classification.
\newblock {\em The Journal of Supercomputing}, 75:5551--5582, 2019.

\bibitem{abbaspour2022comparative}
Akbar Abbaspour Ghadim~Bonab.
\newblock A comparative study of demand forecasting based on machine learning methods with time series approach.
\newblock {\em Journal of applied research on industrial engineering}, 9(3):331--353, 2022.

\bibitem{karimi2022automated}
Mohammad Karimi~Moridani and Atiye Hajiali.
\newblock Automated sleep stage detection based on recurrence quantification analysis using machine learning.
\newblock {\em Journal of applied research on industrial engineering}, 9(4):409--426, 2022.

\bibitem{aftabi2024feed}
Navid Aftabi, Nima Moradi, and Fatemeh Mahroo.
\newblock Feed-forward neural networks as a mixed-integer program.
\newblock {\em arXiv preprint arXiv:2402.06697}, 2024.

\bibitem{rathore2018malware}
Hemant Rathore, Swati Agarwal, Sanjay~K Sahay, and Mohit Sewak.
\newblock Malware detection using machine learning and deep learning.
\newblock In {\em Big Data Analytics: 6th International Conference, BDA 2018, Warangal, India, December 18--21, 2018, Proceedings 6}, pages 402--411. Springer, 2018.

\bibitem{rahul2020analysis}
Rahul, Priyansh Kedia, Subrat Sarangi, and Monika.
\newblock Analysis of machine learning models for malware detection.
\newblock {\em Journal of Discrete Mathematical Sciences and Cryptography}, 23(2):395--407, 2020.

\bibitem{akhtar2023evaluation}
Muhammad~Shoaib Akhtar and Tao Feng.
\newblock Evaluation of machine learning algorithms for malware detection.
\newblock {\em Sensors}, 23(2):946, 2023.

\bibitem{sewak2018comparison}
Mohit Sewak, Sanjay~K Sahay, and Hemant Rathore.
\newblock Comparison of deep learning and the classical machine learning algorithm for the malware detection.
\newblock In {\em 2018 19th IEEE/ACIS international conference on software engineering, artificial intelligence, networking and parallel/distributed computing (SNPD)}, pages 293--296. IEEE, 2018.

\bibitem{liu2017automatic}
Liu Liu, Bao-sheng Wang, Bo~Yu, and Qiu-xi Zhong.
\newblock Automatic malware classification and new malware detection using machine learning.
\newblock {\em Frontiers of Information Technology \& Electronic Engineering}, 18(9):1336--1347, 2017.

\bibitem{hemalatha2021efficient}
Jeyaprakash Hemalatha, S~Abijah Roseline, Subbiah Geetha, Seifedine Kadry, and Robertas Dama{\v{s}}evi{\v{c}}ius.
\newblock An efficient densenet-based deep learning model for malware detection.
\newblock {\em Entropy}, 23(3):344, 2021.

\end{thebibliography}
 
}
\end{document}